\documentclass[10pt]{article}
\usepackage{graphicx}
\usepackage{amsmath}
\usepackage{hanging}
\usepackage{apacite}
\usepackage[lmargin=0.6in, rmargin=0.6in, tmargin=.5in, bmargin=.5in]{geometry}
\usepackage[labelfont=bf]{caption}
\usepackage{authblk}

\usepackage{caption} 
\usepackage{lettrine}
\usepackage{kpfonts}
\usepackage{multicol}
\usepackage{titling}
\usepackage{color}
\usepackage{dblfloatfix}
\usepackage{parnotes}
\makeatletter

\usepackage{booktabs}
\usepackage[flushleft]{threeparttable}

\definecolor{blue}{rgb}{0,0,0.5}

 \pretitle{\begin{flushleft}\LARGE}
 \posttitle{\end{flushleft}}
  \preauthor{\begin{flushleft}}
 \postauthor{\end{flushleft}}
  
  \usepackage{titlesec}
\titlespacing*{\section}{0pt}{10pt}{-1pt}
 \titlespacing*{\subsection}{0pt}{2pt}{-1pt}
 \titlespacing*{\subsubsection}{0pt}{2pt}{-1pt}
\titleformat*{\section}{\normalfont\Large\bfseries\color{blue}}
\titleformat*{\subsection}{\normalfont\large\bfseries\color{blue}}
\titleformat*{\subsubsection}{\normalfont\bfseries\color{blue}}


\bibleftmargin{10pt}
\bibindent{-10pt}

\title{\vspace{-2.0cm}\textbf{\textcolor{blue}{A Network Perspective on Attitude Strength: Testing the Connectivity Hypothesis}}}
\date{}
\author{Jonas Dalege, Denny Borsboom, Frenk van Harreveld \& Han L. J. van der Maas}
\affil{Department of Psychology, University of Amsterdam, 1018 WT Amsterdam, The Netherlands}

\begin{document}

\maketitle

\section*{\vspace{-8ex}} 
\textbf{Attitude strength is a key characteristic of attitudes. Strong attitudes are durable and impactful, while weak attitudes are fluctuating and inconsequential. Recently, the Causal Attitude Network (CAN) model was proposed as a comprehensive measurement model of attitudes, which conceptualizes attitudes  as networks of causally connected evaluative reactions (i.e., beliefs, feelings, and behavior toward an attitude object). Here, we test the central postulate of the CAN model that highly connected attitude networks correspond to strong attitudes. We use data from the American National Election Studies 1980-2012 on attitudes toward presidential candidates (total \textit{n} = 18,795). We first show that political interest predicts  connectivity of attitude networks toward presidential candidates. Second, we show that connectivity is strongly related to two defining features of strong attitudes -- stability of the attitude and the attitude's impact on behavior. We conclude that network theory provides a promising framework to advance the understanding of attitude strength.}

\begin{multicols}{2}

\lettrine[findent=0pt, lines=3]{\textcolor{blue}W}{ }hile some  attitudes are durable and impactful, other  attitudes are largely inconsequential and fluctuating -- in short, attitudes differ in their strength. These fundamental differences between strong and weak attitudes have spurred decennia of research in Political and Social Psychology and several attributes related to attitude strength have been identified (e.g., importance, certainty, accessibility; for overviews see \citeNP{Cunningham2015, Krosnick1995, Visser2006}). Recently, the Causal Attitude Network (CAN) model,  which conceptualizes attitudes as networks of causally connected evaluative reactions (i.e., beliefs, feelings, behaviors toward an attitude object; \citeNP{Dalege2016}), has been proposed as a comprehensive measurement model of attitude. The CAN model holds that connectivity of attitude networks represents a formalized concetptualization of attitude strength. Here, we provide a first of the connectivity hypothesis, which holds that highly connected attitude networks correspond to strong attitudes.
\section*{The Causal Attitude Network (CAN) Model}
The basic premise of the CAN model is that an attitude is a system of evaluative reactions that influence each other \cite{Dalege2016}. Network modeling offers a natural representation of this hypothesis, because in network models, complex systems are modeled as a set of autonomous entities (i.e., nodes) and connections between these nodes (i.e., edges). Together, the set of nodes and edges defines the network structure \cite{Newman2010}. In attitude networks, nodes refer to evaluative reactions such as judging a presidential candidate as competent, charismatic and honest; feeling hope and pride about a presidential candidate; and showing support and voting for a presidential candidate. Edges in attitude networks represent bidirectional pairwise interactions between evaluative reactions (e.g., feeling hope about a presidential candidate may result from judging the candidate as being honest and vice versa). The CAN model is based on recent empirically derived network models as applied to psychopathology, personality, and cognitive psychology \cite<e.g.,>{Cramer2012, Cramer2010, vanderMaas2006} and on theoretical parrelel constraint satifaction models of attitude, which assume that the drive for cognitive consistency is the main force of attitude formation and change \cite<e.g.,>{Monroe2008, Read1997, Shultz1996, Simon2004, Simon2015}. The basic premise of the CAN model -- that attitudes are systems of interconnected beliefs, feelings and behaviors -- is also reminiscent of work by \citeA{McGuire1990} on thought systems. \par
In the CAN model, the average level of the weights depends on how often an individual interacts with the attitude object in one way or another (e.g., thinking about or perceiving the attitude object). Based on the mechanism of constraint satisfaction that was implemented in a recent connectionist model of attitudes (\citeNP{Monroe2008}; see also \citeNP{Cunningham2007} for a related neuroanatomical model), the CAN model proposes that connections between evaluative reactions self-organize when the individual interacts with the attitude object. Attributes related to attitude strength, such as importance \cite{Boninger1995} and elaboration \cite{Petty1995}, make it likely that an individual interacts with the attitude object and that the global connectivity of his or her attitude network increases as a result of this. Politically interested individuals are thus expected to have highly connected political attitudes.\par 
\section*{Global Connectivity of Attitude Networks}
The level of global connectivity is a primary aspect of networks. Global connectivity in networks depends both on the number of connections between nodes and on the magnitude of the connection weights. In highly connected networks, changes in one node have stronger effects on other nodes. Thus, in highly connected attitude networks, evaluative reactions have more causal impact on each other than in weakly connected attitude networks. As a result, nodes in highly connected networks typically align their states, while nodes in weakly connected networks can vary relatively independently and show more random patterns \cite<e.g.,> {Kindermann1980, Scheffer2012}. Next we will illustrate what this means in the context of attitudes, after which we will discuss why connectivity can be regarded as a formalized conceptualization of attitude strength.\par
\begin{figure*}[]
    \centering
\includegraphics[width=6in]{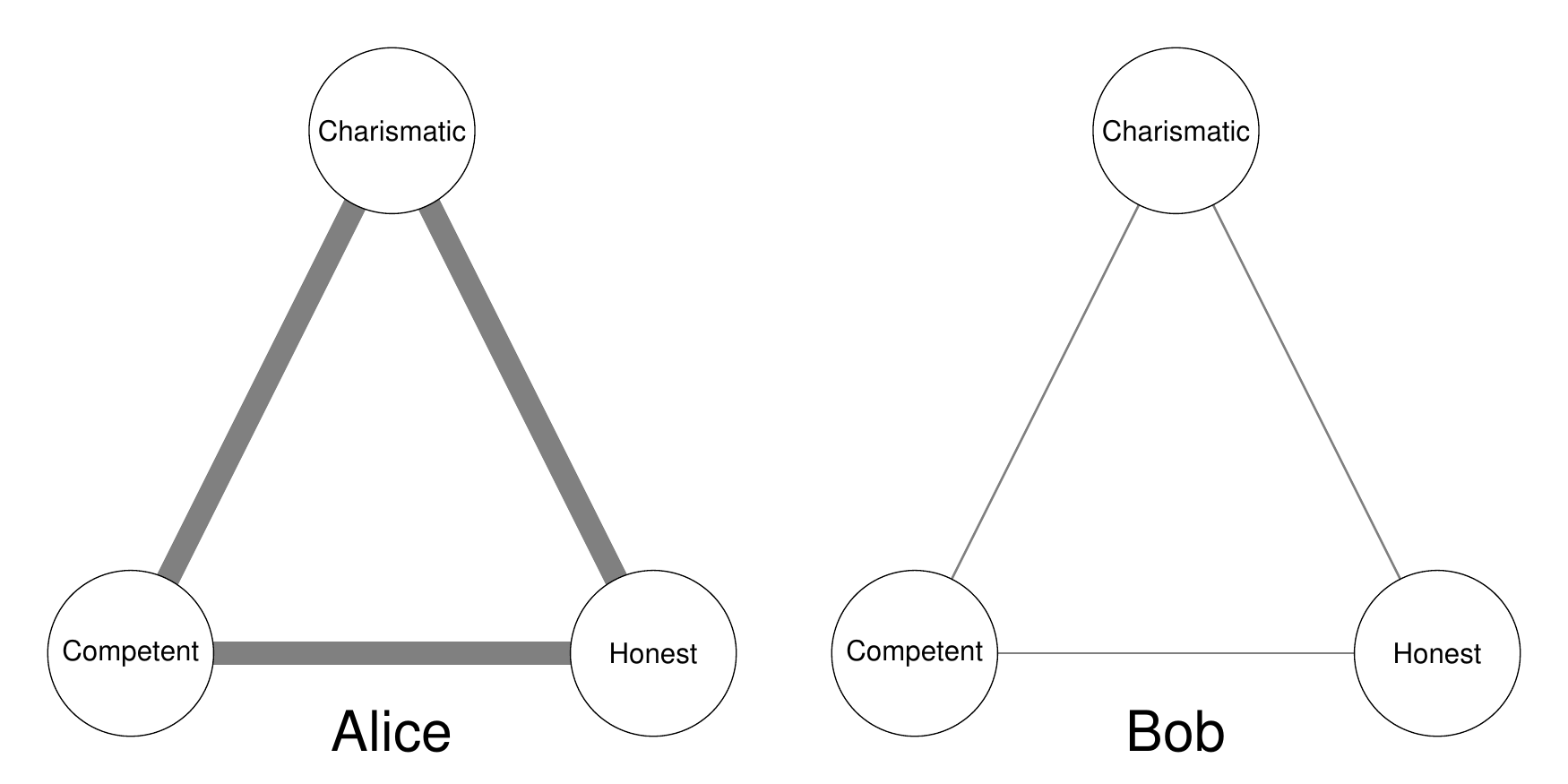}
\caption{Representation of evaluative reactions that are part of a highly connected (Alice) or weakly connected (Bob) attitude network. Thickness of edges represents strength of (reciprocal) causal connections between the evaluative reactions.}
\label{fig:fig1}
\end{figure*}

Consider Alice and Bob, who both have a positive attitude toward Donald Trump. Suppose that both Alice's and Bob's evaluative reactions include judging Donald Trump as charismatic, honest and competent. Their attitudes, however, differ in their connectivity -- Alice's network is highly connected while Bob's network is weakly connected (see Figure \ref{fig:fig1} for a graphic representation). As a result, if Alice were to receive information that is incongruent with the current state of a given evaluative reaction -- say, she learns of a mistake Donald Trump made, implying that Donald Trump might not be as competent as Alice initially thought -- strong causal connections between the judgments would create an unstable state of the network \cite{vanBorkulo2014}. To regain stability, Alice should either discard the information as unreliable or her other judgments should change, too -- the former course of action, however, becomes more likely with increasing connectivity between the judgments up to the point where it becomes extremely difficult to change one of the evaluative reactions in isolation. If Bob, on the other hand, were to change his judgment that Donald Trump is competent, this change would not have much impact on his other judgments, because his network is weakly connected. Thus, conflicting evaluative reactions would disturb his attitude network to a much lesser extent, making it easier to change individual evaluative reactions. Highly connected attitudes thus result in consistent attitudes -- an attribute that is related to attitude strength \cite{Chaiken1995, Eagly1995, Judd1989, Judd1981}. Network connectivity therefore provides a mechanistic explanation of why attitudes differ in their consistency \cite{Dalege2016}.

\section*{Network Connectivity and Attitude Strength}
Dynamical properties of networks, such as their evolution in time and resistance to change, are intimately intertwined with network structure \cite<e.g.,> {Cramer2016, Kolaczyk2009, Manrubia1999, Scheffer2012, Watts2002}. It turns out that the dynamics of highly and weakly connected \textit{networks} bear a striking resemblance to the defining features of strong and weak \textit{attitudes} \cite{Dalege2016}.Note that this implies that connectivity of attitude networks predicts attitude strength. However, the CAN model does not exclude the possibility that also other factors independent of network connectivity predict attitude strength.\par
First of all, strong attitudes are durable -- they are resistant to change and stable \cite<e.g.,> {Bassili1996, Bizer2005, Visser1998}. Highly connected networks are more resistant to change than weakly connected networks. Weakly connected networks change roughly proportional to the external force put on the network's nodes (e.g., a persuasive message regarding Trump's incompetence), while a disproportionate amount of force is needed to instigate change in highly connected networks \cite{Cramer2016}. Furthermore, individual nodes in weakly connected networks will intrinsically show more random variation than nodes in highly connected networks because their behavior is to a larger extent controlled by random perturbations from outside the network \cite{Kindermann1980, vanBorkulo2014}.\par
Second, strong attitudes are better predictors of behavior than weak attitudes \cite<e.g.,> {Bizer2011, Fazio1986, Holland2002}. The reasons that highly connected attitude networks are more likely to guide behavior are threefold. Because the CAN model treats behavior as part of the attitude network, factors that increase connectivity between non-behavioral evaluative reactions (i.e., beliefs and feelings) are expected to also influence the connectivity between non-behavioral evaluative reactions and behavioral evaluative reactions \cite<e.g., behavioral decisions;> {Dalege2017}. This implies that highly connected attitude networks should be more predictive of behavior.\par
Third, strong attitudes exert more influence on information processing -- they direct attention and influence the way in which incoming information is integrated \cite<e.g.,>{Fazio1986, Houston1989, RoskosEwoldsen1992}. While connectivity of attitude networks directly causes the former three features of attitude strength, the influence on information processing in our framework is probably indirect. As discussed above, successful persuasion is more likely to instigate conflict in highly connected networks. It is likely that individuals are motivated to avoid such conflict and are therefore motivated to integrate information in a way that does not disrupt the equilibrium of their attitude network. This motivation also might lead to heightened attention to attitude objects to early detect "attacks" on the attitude network.\par 
Linking network connectivity to attitude strength thus might provide a formalized explanation how the effects of strong attitudes are manifested. However, empirical support for the hypothesis that network connectivity predicts attitude strength is still lacking. The aim of this paper is to provide a first test of this hypothesis and thereby providing more empirical grounding of the CAN model.
\section*{Overview of the Present Research}
To test the connectivity hypothesis, we applied network analysis to the open-access data of the American National Election Studies (ANES) between 1980--2012. These data sets involve large and representative sample sizes and focused on both theoretically and practically highly relevant attitudes -- attitudes toward presidential candidates. Furthermore, the ANES were used in several earlier lines of research into attitude strength and related issues \cite<e.g.,> {Alwin1991, Bizer2004, Crano1997, Eaton2009, Judd1981, Krosnick1988, Krosnick1989b, Lusk1988, Visser2003}.\par
To test the connectivity hypothesis, we first had to identify a variable likely to predict network connectivity. The CAN model assumes that connectivity of attitude networks depends on the amount of interaction with the attitude object and we therefore selected a measure of political interest as the predictor of network connectivity of attitudes toward presidential candidates. Our first analysis tested whether indeed political interest predicts attitudinal network connectivity.\par
While interest is related to attitude strength \cite<e.g.,> {Krosnick1993}, it is not a defining feature of attitude strength. In the second analysis we therefore tested whether network connectivity predicts two defining features of attitude strength. First, we tested whether network connectivity predicts durability of attitudes by assessing attitude stability. Second, we tested whether network connectivity predicts impact of attitudes by assessing the impact of attitudes on behavior. The connectivity hypothesis holds that network connectivity positively predicts attitude strength.
\section*{Method}
\subsection*{Participants}
The ANES samples are large samples that are representative of the US American adult population. Questionnaires were administered in two surveys before and after each American presidential election from 1980 to 2012 by the Center for Political Studies of the University of Michigan. The samples to which all relevant items were administered in the pre-election survey consisted of 18,795 adult participants in total (see Table \ref{tab:tab1} for number of participants per election). This large sample size provides us with high statistical power.

\end{multicols}

\begin{table}[h]

 \begin{threeparttable}
    \caption{Numbers of participants assigned to the interest groups for each election and numbers of participants who had missing values on the interest variable. Numbers of excluded participants are shown in parentheses for the attitudes toward democratic and republican candidates, respectively.}
      \label{tab:tab1}
 \begin{tabular}{lllll}
        \toprule
        Election & Low interest groups & Intermediate interest groups & High interest groups & Missings interest \\
        \midrule
                 
        1980 & 407 (121, 149) & 692 (78, 100) & 407 (50, 16) & 49   \\
        1984 & 558 (132, 123) &	1054 (123, 170) &	638 (53, 81) &	7 \\
        1988	& 509 (139, 134) &	960 (102, 133) &	567 (50, 57) &	4\\
	1992	 & 419 (89, 60)	& 635 (101, 64) &	304 (48, 20) &	1\\
        1996	& 399 (38, 87)	& 848 (31, 100)	 & 467 (22, 33) &	0\\
	2000 &	396 (137, 111) &	886 (148, 131) &	525 (76, 44) &	0\\
	2004	 & 186 (71, 31) &	528 (102, 32) & 	498 (57, 20) &	0\\
	2008 &	378 (72, 87) &	815 (89, 93) &	1128 (88, 97) &	1\\
	2012	 & 895 (73, 121) &	2460 (68, 153) &	2554 (43, 81) &	 5\\
        \bottomrule
     \end{tabular}
    \begin{tablenotes}
      \item\textit{Note.} A subsample of the ANES 1996 already participated in the ANES 1992. The number of participants who completed the pre-election survey of the ANES 1996 and also participated in the ANES 1992 is 1,316.
    \end{tablenotes}
  \end{threeparttable}
\end{table}

\begin{multicols}{2}

\subsection*{Measures}
Relevant measures included non-behavioral evaluative reactions toward presidential candidates, a measurement of the interest participants had in the campaign, a measure of the global attitude toward presidential candidates, assessed before and after each election, and an assessment of whom the participants voted for.
\subsubsection*{Evaluative Reactions}
Six to 16 items tapping beliefs regarding the presidential candidates and between four to eight items tapping feelings toward the presidential candidates were assessed in the pre-election surveys of the different studies. Note that some items were administered in every ANES or in most ANES, while other items were only administered infrequently, see Table \ref{tab:tab2}.\par
For items tapping beliefs, participants were asked: "In your opinion, does the phrase 'he...' describe \textit{the candidate}...?" and Table 2 shows the phrases that completed the items. In the ANES from 1980 to 2004, and for a subsample of the ANES of 2008 (N = 1133), items were assessed on a 4-point scale, with answer options 4 = "Extremely well", 3 = "Quite well", 2 = "Not too well", 1 = "Not well at all". In order to use current network estimation software \cite{vanBorkulo2014}, we dichotomized the data into two categories, one consisting of responses 1 and 2, and one consisting of responses 3 and 4. For a subsample of participants of the ANES of 2008 (N = 1133), and for all participants in the ANES of 2012, the items were assessed on a 5-point scale, with the answer options 5 = "Extremely well", 4 = "Quite well", 3 = "Moderately well", 2 = "Slightly well", 1 = "Not well at all". Here, we assigned option 1, 2 and 3 in one category and option 4 and 5 in the other category.\footnote{As a check on the robustness of results, we also ran the analyses with answer option "moderately well" assigned to the second category and without dichotomization (by using polychoric correlations).  The results of these alternative analyses mirrored the results reported in this paper.}\par
For items tapping feelings, participants were asked: "Has \textit{the candidate} -- because of the kind of person he is or because of something he has done, ever made you feel: ...?" and the feelings that completed the items are shown in Table 1. These items were assessed dichotomously, with the answer options 1 = "Yes" and 0 = "No".
\subsubsection*{Political Interest}
Political interest was assessed by the item: "Some people don't pay much attention to political campaigns. How about you? Would you say that you have been... in the political campaigns so far this year?"  We assigned participants who answered "very much interested", "somewhat interested", "not much interested" to high, intermediate, weak interest groups, respectively.\footnote{In the ANES of 2008, the item was only administered to a subsample (\textit{N} = 1178), while another subsample (\textit{N} = 1144) was administered the item: "How interested are you in information about what's going on in government and politics?" The answer options were "Extremely interested", "Very interested", "Moderately interested", "Slightly interested", and "Not interested at all". Here, we assigned participants who answered either "Extremely interested" or "Very interested", "Moderately interested", "Slightly interested" or "Not interested at all" to high, intermediate, weak attitude strength groups, respectively. The frequencies of the number of participants assigned to each group were similar for both interest items.}   The number of participants assigned to each strength group at each election can be found in Table \ref{tab:tab1}.
\subsubsection*{Global Attitude Measure} 
In both the pre- and post-election interview, participants rated how warm and favorable they felt toward the presidential candidate on a scale from 0 to 100, with 0 representing a very unfavorable attitude toward the presidential candidate and with 100 representing a very favorable attitude toward the presidential candidate.    
\subsubsection*{Voting Decision} 
In the post-election interview, participants were asked which candidate they voted for. Depending on which presidential candidate the analysis focused, we scored the response as 1 when the participants stated that they voted for the given candidate and we scored the response as 0 when the participants did not vote for the given candidate \cite<cf.,> {Payne2010}. 
\subsection*{Data Analysis}
To estimate attitude network structures, we applied the eLasso-procedure \cite{vanBorkulo2014} to the responses on the non-behavioral evaluative reactions.\footnote{We did not include voting behavior in the estimation of the attitude networks in this analysis because (a) we wanted to use predictability of voting behavior as an attitude strength index and (b) voting behavior was not assessed at the same time as the other evaluative reactions.}  In the eLasso-procedure, each variable is regressed on all other variables, while the regression function is subjected to regularization to control the size of the statistical problem \cite<see> {Friedman2008, Tibshirani1996}. For each node, the set of edges that displays the best fit to the data is selected based on the fit of the regression functions according to the Extended Bayesian Information Criterion \cite{Chen2008}. Parameters are then based on the regression parameters in the selected neighborhood functions \cite{vanBorkulo2014}. For each interest group at each election, we estimated networks for the non-behavioral evaluative reactions toward each candidate by using this procedure, which resulted in a total of 54 estimated networks. For further information on estimating and analyzing attitude networks, see \citeA{Dalege2017c}.\par
We used the Average Shortest Path Length \cite<ASPL;> {West1996} as a measure of network connectivity. For every two given nodes in the network, the ASPL computes the length of the shortest path that connects the nodes and then averages these estimates. Dijkstra's algorithm was used to calculate path lengths \cite{Brandes2001, Dijkstra1959, Newman2001}. Dijkstra's algorithm minimizes the inverse of the distance between two node pairs using the weight of the edges (the absolute values of the regression parameters in the networks reported in this article). A low ASPL indicates high connectivity, while a high ASPL indicates low connectivity.\par
To investigate whether political interest predicts network connectivity, we  fitted a general linear model with the interest groups as factor, the ASPL as dependent variable, and the number of nodes in the network as covariate to control for network size. For pairwise comparisons, we used Tukey's test that corrects p-values for multiple testing.\par

\end{multicols}

\begin{table}[t]
\begin{threeparttable}
    \caption{Included and excluded evaluative reactions.}
    \label{tab:tab2}
   \begin{tabular}{ll}
        \toprule
        Evaluative Reaction &	Included (Excluded)\\
        \midrule
        \multicolumn{2}{l}{\underline{Items tapping beliefs}}\\       
"really cares about people like you" &	1984--2012 ($1984^{D}$)\\
"is compassionate" &	1984--1996* ($1984^{D}$, 1988)\\
"is decent" &	1984, 1988 ($1988^{D}$)\\
"is dishonest" &	1980, 2000, 2004\\
"would solve our economic problems" &	1980\\
"gets things done" &	1992, 1996 ($1992^{D}$)\\
"sets a good example" &	1984 ($1984^{D}$)\\
"is fair" &	1984 ($1984^{D}$)\\
"would develop good relations with other countries" &	1980\\
"is hard-working" &	1984 ($1984^{D}$)\\
"is honest" &	1988--1996, 2008, 2012 ($1988^{R}$, $1992^{D}$)\\
"is in touch with ordinary people" & 	1984 ($1984^{D}$)\\
"is inspiring" &	1980, 1984, 1992, 1996 ($1984^{D}$)\\
"is intelligent" &	1984--2012*\\
"can't make up his mind" &	2004\\
"is kind" &	1984	 ($1984^{D}$)\\
"is knowledgeable" &	1980--2012 ($1988^{D}$)\\
"would provide strong leadership" &	1980--2012 ($1984^{D}$,  $1992^{D}$))\\
"is moral" &	1980--2012 ($1984^{D}$, 1988, ($2000^{R}$))\\
"is optimistic" &	2008\\
"is out of touch with ordinary people" &	2000 \\
"is religious" &	1984 (1984)\\
"commands respect" &	1984 ($1984^{D}$)\\
"understands people like you" &	1984 ($1984^{D}$\\
"is power-hungry" &	1980\\
"is weak" &	1980 ($1980^{R}$, $2000^{R}$)\\
         \multicolumn{2}{l}{\underline{Items tapping feelings}}\\       
"angry" &	1980--2012 ($1980^{R}$)\\
"afraid of him" &	1980--2012 ($1980^{R}$\\
"disgusted" &	1980, 1984\\
"hopeful" &	1980--2012\\
"proud" &	1980--2012\\
"sympathetic toward him" &	1980, 1984\\	
"uneasy" &	1980, 1984\\
        \bottomrule
     \end{tabular}
    \begin{tablenotes}
      \item\textit{Note.} *In 1992, these items were only assessed for Bill Clinton, $^{D}$ the item was only excluded for the Democratic candidate, $^{R}$ the item was only excluded for the Republican candidate.
    \end{tablenotes}
  \end{threeparttable}
\end{table}

\begin{multicols}{2}

As a measure of the attitude's impact on behavior, we calculated the biserial correlation between the global attitude measure assessed before the election and the voting decision. We estimated the attitude's impact on behavior for each interest group at each election and for each candidate. To test whether network connectivity predicts the attitude's impact on behavior we calculated the partial correlation between network connectivity and the attitude's impact on behavior on the interest group level with network size partialled out.\par
As a measure of the attitude's stability, we calculated the Pearson correlation between the global attitude measure assessed before and after the election. We estimated the attitude's stability for each interest group at each election and for each candidate. To test whether network connectivity predicts the attitude's stability we calculated the partial correlation between network connectivity and the attitude's stability on the interest group level with network size partialled out.\par
The vast majority of missing values in the pre-election interview were caused by participants answering "Don't know" on items tapping evaluative reactions. We omitted variables that had more than 10\% missing values to decrease the number of participants who had to be excluded, as we applied casewise deletion to cases with missing values.\footnote{We also ran the analyses with inclusion of all variables, imputed missing values randomly with either 0 or 1 or scored "Don't know" as a middle point of the scale, controlled for sample size and variance, and estimated networks using polychoric correlations. The results of these analyses mirrored the results reported in this paper.}  The numbers of these excluded participants can be found in \ref{tab:tab1} and Table \ref{tab:tab2} shows which variables were omitted. Participants, who did not answer the post-election interview or who did not vote in the presidential election had to be excluded for the analyses on the attitude's impact on behavior and stability (we excluded participants only for the relevant analyses). 
\subsection*{Results}
\begin{figure*}[]
    \centering
\includegraphics[width=6in]{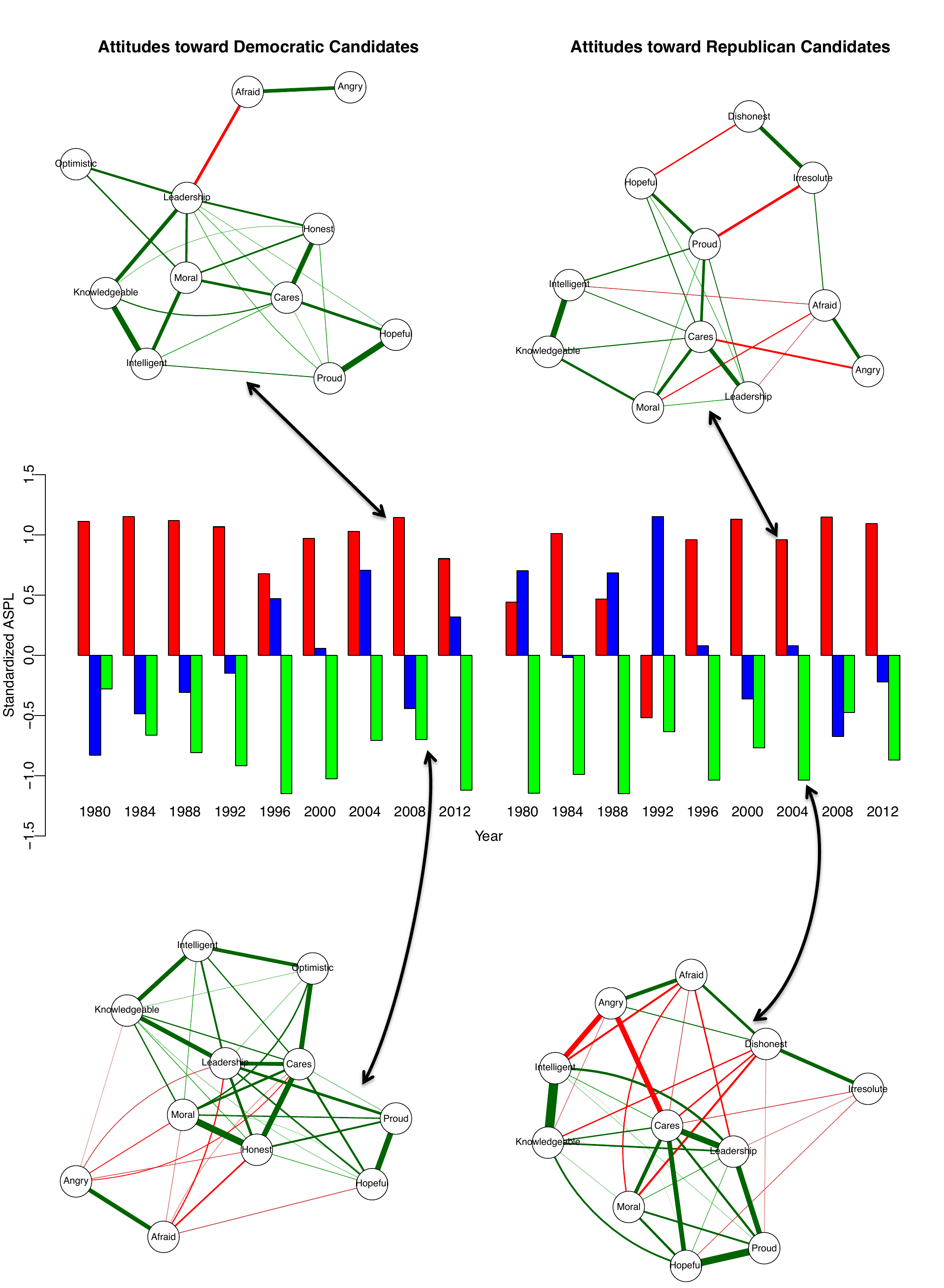}
\caption{The barplot shows the standardized ASPL (number of standard deviations above or below mean for each set of interest groups) for each candidate at each year and for each interest group. Red (blue) [green] bars represent low (intermediate) [high] interest groups. Two representative networks of the low (high) interest groups are shown above (below) the barplot. The left (right) networks represent attitude networks toward Barack Obama (George W. Bush) in 2008 (2004). Nodes represent evaluative reactions (see Table \ref{tab:tab2} for the complete wording of the items), green lines represent positive connections, red lines represent negative connections, and thickness of an edge represents the strength of the connection. Closely connected nodes are placed near each other \protect\cite{Fruchterman1991}. All networks shown in this paper were constructed using the R-package qgraph \protect\cite{Epskamp2012}.}

\label{fig:fig2}
\end{figure*}

Political interest strongly predicted connectivity of attitude networks (see Figure \ref{fig:fig2}). Note that some of the low interest networks (Mondale 1984, Bush 1988, Kerry 2004) were not fully connected (i.e. not all nodes were directly or indirectly connected), which leads to infinitely large shortest path lengths between disconnected nodes. To be able to still enter such networks into the analysis, we set infinitely large shortest path lengths to the highest shortest path length in the same network that was a real number. Note that this technique results in overestimation rather than underestimation of a network's connectivity, which implies that this technique is conservative with respect to the focal hypothesis (i.e., leads to higher connectivity estimates in low interest groups).\par

The fitted linear model (including the number of nodes as a covariate) showed a significant effect of the interest groups on the ASPL of the networks, \textit{F} (2, 50) = 17.59, \textit{p} $<$ .001, \textit{$\eta_p^2$} = .41. All groups differed significantly from each other in the expected direction and these differences were marked by high effect sizes. The mean ASPL of the intermediate interest groups (\textit{M} = 2.07) was lower than the mean ASPL of the low interest groups (\textit{M} = 2.44), \textit{t} (50) = 3.43, \textit{p} = .004, 95\% CI [0.16; 0.58], \textit{d} = 1.62, indicating that the networks of the intermediate interest groups had a higher connectivity than the networks of the low interest groups. The mean ASPL of the high interest groups (\textit{M} = 1.80) was lower than the mean ASPL of the intermediate interest groups, \textit{t} (50) = 2.48, \textit{p} = .044, 95\% CI [0.06; 0.48], \textit{d} = 1.17, indicating that the networks of the high interest groups had a higher connectivity than the networks of the intermediate interest groups. The mean ASPL of the high interest groups was lower than the mean ASPL of the low interest groups, \textit{t} (50) = 5.91, \textit{p} $<$ .001, 95\% CI [0.43; 0.85], \textit{d} = 2.78, indicating that the networks of the high interest groups had a higher connectivity than the networks of the low interest groups.\par
\begin{figure*}[]
    \centering
\includegraphics[width=7.5in]{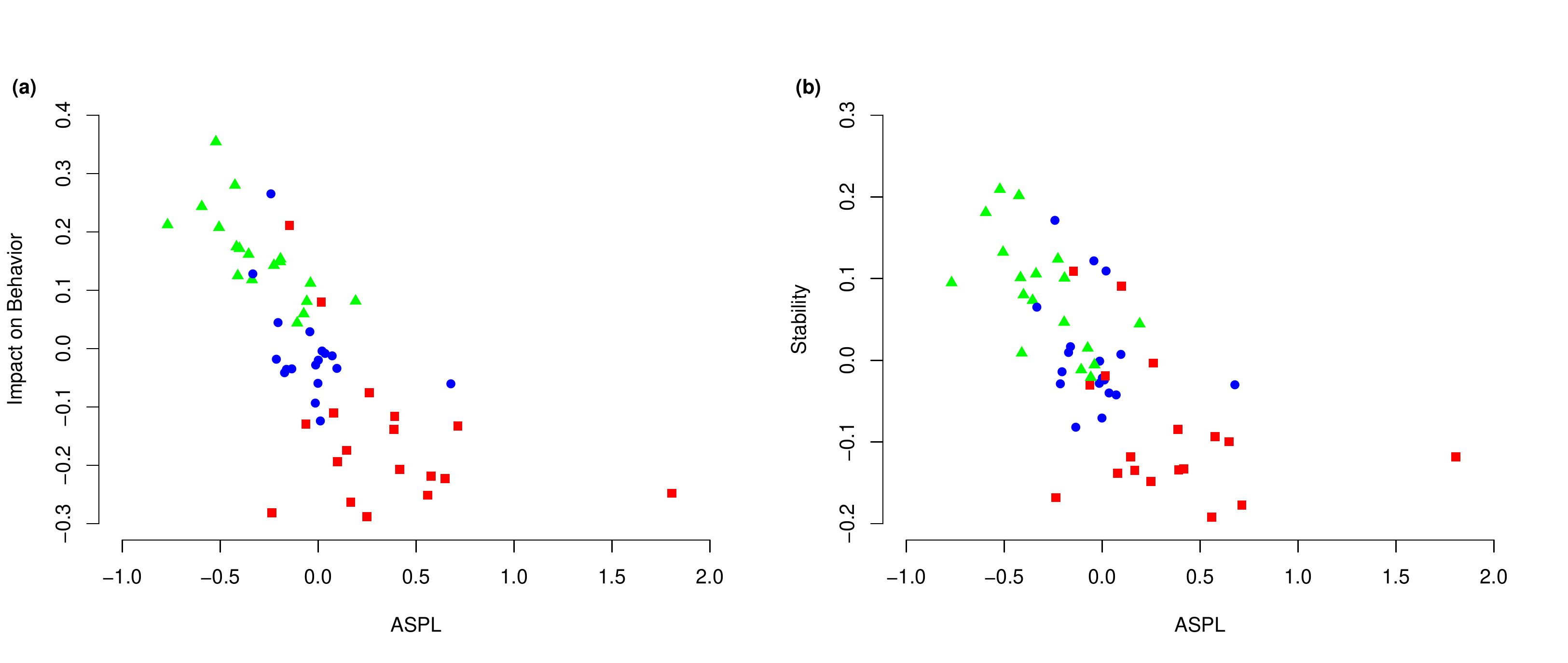}
\caption{Relation between network connectivity (ASPL) and attitude strength (impact on behavior and stability). (a) Scatterplot of the relation between ASPL and attitude's impact on behavior (controlled for network size). (b) Scatterplot of the relation between ASPL and attitude stability (controlled for network size). Red squares (blue circles) [green traingles] represent low (intermediate) [high] interest groups. See the online article for the color version of this figure.}
\label{fig:fig3}
\end{figure*}

The finding that political interest predicts connectivity of attitude networks allowed us to test the central hypothesis of this paper -- that network connectivity predicts attitude strength. This hypothesis was strongly supported. Network connectivity (after controlling for the size of the network) was highly related to both the attitude's impact on behavior, \textit{r} = -.71, \textit{p} $<$ .001, see Figure \ref{fig:fig3}a, and to the attitude's stability, \textit{r} = -66, \textit{p} $<$ .001, see Figure \ref{fig:fig3}b. This indicates that network connectivity is related to both the durability and impact of attitudes.

\section*{Discussion}
We applied network analysis to the archival data of the ANES 1980-2012. By focusing on attitudes toward presidential candidates, we investigated the central postulate of the Causal Attitude Network (CAN) model that strong attitudes correspond to highly connected attitude networks. We first showed that political interest strongly predicts network connectivity of attitudes. Because interest in attitude objects is related to but not a defining feature of attitude strength, we then tested whether connectivity of attitude networks predicts two central features of attitude strength -- the attitude's impact on behavior and stability. Network connectivity strongly predicted both impact on behavior and stability, supporting the central postulate of the CAN model that highly connected attitude networks correspond to strong attitudes. 
\subsection*{Limitations}
While using archival data from the ANES for our analyses has several advantages (e.g., large and representative sample sizes), it also has some drawbacks. The measure of interest had a conceptually different focus (i.e., political campaigns) than the attitude measures (i.e., presidential candidates). However, a measure of interest more closely related to the measures of attitude is more likely to increase the relation between interest and network connectivity than to decrease it. We would also argue that during the election year, political campaigns are focused on the presidential candidates -- making the measure of interest more closely related to the attitude measures.\par 
Another limitation of the current study is that it only included measures of two of the four central features of attitude strength. Feature research therefore should investigate the relation between network connectivity and (1) resistance to persuasion and (2) impact on information processing. Additionally, the current study only measured one specific set of attitudes -- attitudes toward presidential candidates. Future research might investigate the generality of the relation between network connectivity and attitude strength by focusing on a more diverse set of attitudes.\par
Given that the estimated networks were based on groups of individuals, it is not straightforward to generalize our findings to the level of the individual \cite<e.g.,> {Borsboom2013}. However, the group-based networks reported in this paper are likely to be representative of individually based networks if the groups are relatively homogenous \cite<e.g., individuals belonging to the same group have similar network structures;> {vanBorkulo2015}. It is likely that groups were homogenous because the assignment to political interest groups made the groups more homogenous and it is also likely that connections between different evaluative reactions differ more in quantity than in quality (e.g., the connection between judging a presidential candidate to be caring and judging a presidential candidate to be moral probably is positive for most individuals). Nonetheless, future research should investigate whether our findings also replicate at the individual level using appropriate techniques. The challenge here is that attitude networks currently can only be estimated based on several data points (either several data points based on several persons or based on several measurements per person). This makes it especially difficult for research at the individual level, because a highly connected network might limit the variation at the individual level, so that strong connections might not be recovered because of too low variation. 
\subsection*{Implications and Directions for Future Research}
Our results suggest that at least some strong attitudes are based on highly connected networks. This finding has fundamental implications for theorizing on attitude strength and attitude-behavior consistency. Linking attitude strength to network connectivity provides a novel and promising way to derive predictions regarding the dynamics of strong attitudes (that are based on highly connected networks). Such attitudes will generally be highly stable, but can also show instances of high instability under specific circumstances. This would be the case when the attitude network is highly connected but at the same time has to integrate a large amount of conflicting information.\par
Network connectivity also provides hypotheses regarding \textit{how} attitudes change  \cite{Dalege2016}. Change in weakly connected attitude networks takes place on a continuum ranging from a configuration in which all evaluative reactions are negative, to a configuration in which all evaluative reactions are positive, with unaligned configurations being only slightly less stable than aligned configurations. Change in highly connected attitude networks, in contrast, occurs more in an all-or-none fashion, with aligned configurations being much more stable than unaligned configurations. These dynamical characteristics of weakly versus highly connected attitude networks link our work to the catastrophe model of attitudes \cite{Latane1994, vanderMaas2003, Zeeman1976}, which holds important attitudes act like categories (i.e., attitudes can be either positive or negative), while unimportant attitudes act like dimensions (i.e., attitudes represent a continuous dimension running from positive to negative). Linking the connectivity hypothesis to the catastrophe model of attitudes leads to the prediction that important (unimportant) attitudes act like categories (dimensions) because they correspond to highly (weakly) connected networks.\par
Focusing on the connectivity of attitude networks provides novel opportunities for both predicting and influencing (voting) behavior \cite<see also> {Dalege2017}. Regarding behavior prediction, identifying whether individuals' attitude networks are densely or sparsely connected can inform researchers whether it is likely or not that the measured evaluative reactions will predict subsequent behaviors. Pollsters might befit from that because this provides a novel opportunity to estimate how likely it is that the results of a given poll will translate into election results. Regarding influencing behavior, connectivity of attitude networks are can inform which strategy should be taken to influence behavior. If the attitude network is weakly connected, a focus on the behavior itself may be a means to influence behavior effectively. If the attitude network is highly connected, however, the most promising way to influence behavior is probably to first apply strategies to decrease the connectivity of the attitude network (e.g., by lowering the importance of the attitude). When the connectivity has decreased, it is probably easier to induce the desirable behavior. To enhance longevity of attitude change, it would be beneficial to heighten the connectivity of the attitude network after the desired behavior is induced.\par
An interesting avenue for future research is also to investigate how network connectivity relates to other attributes of attitude strength, such as ambivalence, certainty or accessibility. While it is beyond the scope of the current article to discuss the relation between network connectivity and all attributes related to strength, we discuss some potentially fruitful avenues for future research on such relations. First, based on the CAN model \cite{Dalege2016}, we would argue that it is likely that network connectivity amplifies the relation between potential ambivalence (i.e., number of conflicting evaluations) and felt ambivalence (i.e., discomfort resulting from ambivalence, e.g., \citeNP{Priester1996}). Second, knowledge on the attitude object was shown to amplify the effects of attitude strength (e.g., high knowledge in combination with high attitude strength leads to very pronounced stability and impact of attitudes). Similarly, network size (which represents a straightforward conceptualization of knowledge on the attitude object) amplifies the effects of network connectivity. For a more comprehensive discussion of the relations between network connectivity and attitude strength, we refer the interested reader to \citeA{Dalege2016}.
\subsection*{Conclusion}
In this paper, we provided support for the connectivity hypothesis, which holds that highly connected attitude networks correspond to strong attitudes. The connectivity hypothesis provides several novel pathways for research into the different dynamics of strong and weak attitudes, such as instances of high instability of strong attitudes, indicating that network theory shows promise in providing a novel and fruitful framework of attitude strength.

\bibliography{bib} 
\bibliographystyle{apacite}

\subsubsection*{Acknowledgements} 
We thank J. Degner, S. Epskamp, and L. J. Waldorp for comments and discussion.. D. B. was supported by a Consolidator Grant No. 647209 from the European Research Council.
\subsubsection*{Author Contributions} 
J.D. developed the study concept; J.D., D.B., F.v.H., and H.L.J.v.d.M contributed to the study design; J. D. performed the data analysis and interpretation under the supervision of D.B., F.v.H., and H.L.J.v.d.M.; J.D. drafted the manuscript, and D.B., F.v.H., and H.L.J.v.d.M. provided critical revisions.\subsubsection*{Author Information} 
Data used in this paper is available at www.electionstudies.org. Correspondence and request for materials should be addressed to J.D. (j.dalege@uva.nl).
\subsubsection*{Competing Financial Interest} 
The authors declare no competing financial interests.

\end{multicols}

\end{document}